# Power Saving Evaluation with Automatic Offloading

Yoji Yamato[a,*]

[a]Network Service Systems Laboratories, NTT Corporation, 3-9-11 Midori-cho, Musashino-shi, Tokyo 180-8585, Japan

*Corresponding Author: yoji.yamato.wa@hco.ntt.co.jp

## Abstract

Heterogeneous hardware other than small-core CPU such as GPU, FPGA, or many-core CPU is increasingly being used. However, heterogeneous hardware usage presents high technical skill barriers such as familiarity with CUDA. To overcome this challenge, I previously proposed environment-adaptive software that enables automatic conversion, automatic configuration, and high-performance and low-power operation of once-written code, in accordance with the hardware to be placed. I also previously verified performance improvement of automatic GPU and FPGA offloading. In this paper, I verify low-power operation with environment adaptation by evaluating power utilization after automatic offloading. I compare Watt*seconds of existing applications after automatic offloading with the case of CPU-only processing.

**Keywords**: Environment-Adaptive Software, FPGA, Automatic Offloading, Low Power, Evolutionary Computation.

## 1. Introduction

As Moore's Law slows down, a central processing unit's (CPU's) transistor density cannot be expected to double every 1.5 years. To compensate for this, more systems are using heterogeneous hardware, such as graphics processing units (GPUs), field-programmable gate arrays (FPGAs), and so on. For example, Microsoft's search engine Bing uses FPGAs (1), and Amazon Web Services (AWS) provides GPU and FPGA instances (2) using cloud technologies (3)-(10). Systems with Internet of Things (IoT) devices are also increasing (11)-(20).

However, to properly utilize devices other than small-core CPUs in these systems, configurations and programs must be made that consider device characteristics, such as Open Multi-Processing (OpenMP) (21), Open Computing Language (OpenCL) (22), and Compute Unified Device Architecture (CUDA) (23). In addition, programmers need to be sufficiently skilled at using embedded software to precisely control IoT devices. Therefore, for most programmers, the skill barriers are high.

In short, the expectations for applications using heterogeneous hardware are becoming higher, but the skill hurdles for using them are currently high. To surmount these hurdles, application programmers should only need to write logics to be processed, and then software should adapt to the environments with heterogeneous hardware to make it easy to use such hardware.

Java (24), which appeared in 1995, caused a paradigm shift in environment adaptation that enables software written once to run on another CPU machine. However, no consideration was given to the application performance and power consumption at the porting destination. Therefore, I previously proposed environment-adaptive software that effectively runs once-written applications with high performances and low power by automatically converting and configuring code so that GPUs, FPGAs, many-core CPUs, and so on can be appropriately used in deployment environments. For an elemental technology for environment-adaptive software, I also proposed a method for automatically offloading loop statements and function blocks of applications to GPUs or FPGAs and improved performances (25)(26)(27). However, only processing time after offloading has been evaluated, and power consumption after offloading has not been evaluated so far.

This paper proposes a method that takes power consumption into consideration when offloading a normal CPU program to a device such as an FPGA to improve its performance, and verifies the reduction of power consumption after offloading of existing applications. I also propose a method to automatically select an appropriate offload destination in consideration of the power

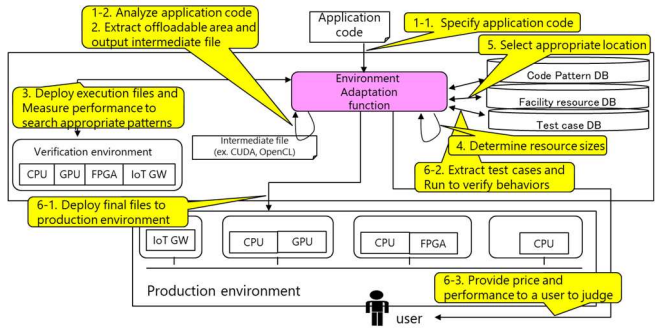

Fig. 1. Processing flow of environment-adaptive software

consumption when offloading is tried in a mixed environment such as a GPU and FPGA.

The rest of this paper is organized as follows. In Section 2, I review technologies on the market and our previous proposals. In Section 3, I propose an automatic offloading method for GPUs and FPGAs that takes power consumption into consideration. In Section 4, I evaluate the power consumption during automatic FPGA offload with the existing application. In Section 5, I conclude the paper.

## 2. Existing Technologies

### 2.1 Technologies on the market

Java is one example of environment-adaptive software. In Java, by using a virtual execution environment called Java Virtual Machine, written software can run even on machines that use different operating systems (OSes) without more compiling (Write Once, Run Anywhere). However, whether the expected performance could be attained at the porting destination was not considered, and too much effort was involved in performance tuning and debugging at the porting destination (Write Once, Debug Everywhere).

CUDA is a major development environment for general-purpose GPUs (GPGPUs (28) that use GPU computational power for more than just graphics processing. To control heterogeneous hardware such as FPGA and GPU uniformly, the OpenCL specification and its software development kit (SDK) are widely used (29). CUDA and OpenCL require not only C language extension but also additional descriptions such as memory copy between devices and CPUs. Because of these programming difficulties, there are few CUDA and OpenCL programmers.

For easy heterogeneous hardware programming, there are technologies that specify parallel processing areas by specified directives, and compilers transform these specified parts into device-oriented codes on the basis of the specified directives. Open accelerators (OpenACC) (30) and OpenMP are examples of directive-based specifications, and the Portland Group Inc. (PGI) compiler (31) and gcc are examples of compilers that support these directives.

In this way, CUDA, OpenCL, OpenACC, OpenMP, and others support GPU, FPGA, or many-core CPU offload processing. Although processing on devices can be done, sufficient application performance and power reduction are difficult to attain. For example, when users use an automatic parallelization technology, such as the Intel compiler (32) for multi-core CPUs, possible areas of parallel processing such as "for" loop statements are extracted. However, naive parallel execution performances with devices are not high because of overheads of CPU and device memory data transfer. To achieve high application performance with devices, CUDA, OpenCL, or so on need to be tuned by highly skilled programmers, or an appropriate offloading area needs to be searched for by using the OpenACC compiler or other technologies.

Therefore, users without skills in using GPU, FPGA, or many-core CPU will have difficulty attaining high application performance and power reduction. Moreover, if users use automatic parallelization technologies to obtain high performance, much effort is needed to determine whether each loop statement is parallelized.

### 2.2 Previous proposals

On the basis of the above background, to adapt software to an environment, I previously proposed environment-adaptive software (26), the processing flow of which is shown in Figure 1. The environment-adaptive software is achieved with an environment-adaptation function, test-case database (DB), code-pattern DB, facility-resource DB, verification environment, and production environment.

Step 1: Code analysis
Step 2: Offloadable-part extraction
Step 3: Search for suitable offload parts
Step 4: Resource-amount adjustment
Step 5: Placement-location adjustment
Step 6: Execution-file placement and operation verification
Step 7: In-operation reconfiguration

Because most offloading to heterogeneous devices is currently done manually, I proposed the concept of environment-adaptive software and automatic offloading to heterogeneous devices. For automation, I also have proposed a method using evolutionary computation to search for appropriate parallel processing parts when offloading to a GPU. However, the previous paper only evaluated the shortening of the processing time and not the reduction of power consumption. Therefore, in this paper, I evaluate the reduction of power consumption when automatically offloading to an offload device such as FPGA.

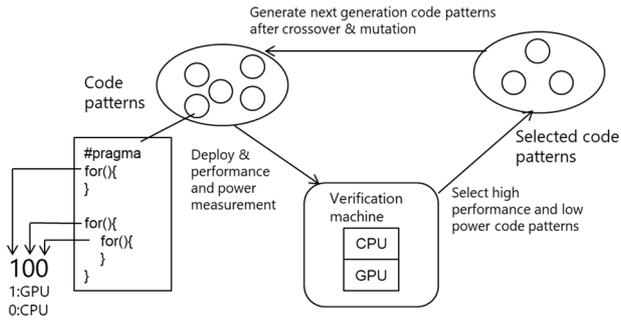

Fig. 2. Automatic GPU offload method considering power consumption

## 3. Automatic GPU and FPGA Offload Considering Power Consumption

To embody the concept of environment-adaptive software, I have so far proposed automatic GPU and FPGA offload of program loop statements, automatic offload of program functional blocks, and multilingual and mixed environments offload. Based on these elemental technologies, in subsections 3.1 and 3.2, I propose automatic GPU and FPGA offload technology for loop statements that take power consumption into consideration. In 3.3, I propose an appropriate offload destination selection technology in a mixed environment of migration destinations.

### 3.1 Automatic GPU offload of loop statements

For automatic GPU offloading of loop statements, I proposed a method and evaluated processing time improvement (33).

First, as a basic problem, the compiler can find the limitation that this loop statement cannot be processed in parallel on the GPU, but it is difficult to find out whether this loop statement is suitable for parallel processing on the GPU. Loop statements with a large number of loops are generally said to be more suitable, but it is difficult to predict the performance and power consumption by offloading to the GPU without actually measuring them. Therefore, it is often the case that the instruction to offload this loop to the GPU is manually given and the performance measurement is tried. On the basis of that, (33) proposed automatically finding an appropriate loop statement that is offloaded to the GPU with a genetic algorithm (GA) (34), which is an evolutionary computation method. From a general-purpose program for normal CPUs, the proposed method first checks the parallelizable loop statements. Then for the parallelizable loop statements, it sets 1 for GPU execution and 0 for CPU execution. The value is set and geneticized, and the performance verification trial is repeated in the verification environment to search for an appropriate area. Here, the pattern that can be processed in a short time in the verification environment measurement is regarded as a gene with high goodness of fit. In this paper, the power consumption is also measured in the verification environment measurement, and a new process is added to make the high goodness of fit for the low power consumption pattern. For example, (Processing time)$^{-1/2}$*(Power consumption)$^{-1/2}$ is set to increase goodness of fit value for short processing time and low power consumption (Figure 2).

(33) also proposed a method for transferring variables efficiently. Regarding the variables used in the nested loop statement, when the loop statement is offloaded to the GPU, the variables that have no problems even if CPU-GPU transfer is performed at the upper level are summarized at the upper level. In addition, for not only nesting but also variables defined in multiple files, GPU processing and CPU processing are not nested, and for variables where CPU processing and GPU processing are separated, the proposed method specifies to transfer them in a batch.

In summary, I propose an evolutionary computation method that includes power consumption in the goodness of fit and a reduction in CPU-GPU transfer. By using them, the speed is increased and the power is reduced automatically.

### 3.2 Automatic FPGA offload of loop statements

I have also proposed a method for offloading loop statements to FPGA to improve performance (26).

When considering offloading a specific loop statement that takes a long time to speed up to FPGA, it is difficult to predict which loop should be offloaded to speed up. Therefore, it is proposed that the performance measurement be performed automatically in the verification environment similar to GPU offload case. However, since it takes several hours or more for an FPGA to compile OpenCL and operate it on an actual machine, it takes a huge amount of processing time to repeatedly measure the performance using GA like automatic GPU offload.

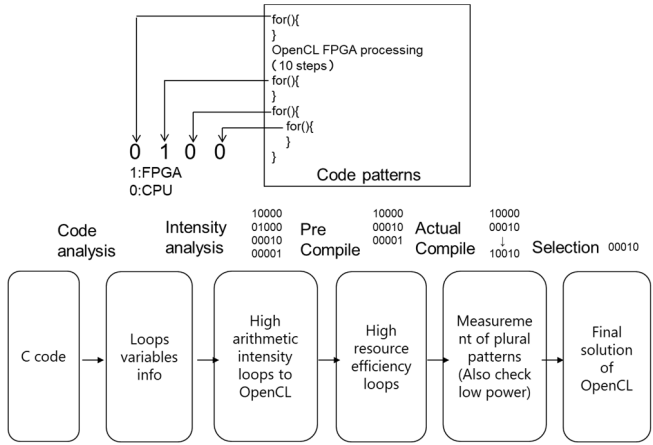

Fig. 3. Automatic FPGA offload method considering power consumption

Therefore, after narrowing down the candidate loop statements to be offloaded to the FPGA, the performance measurement trial is performed. Specifically, first, for the found loop statement, a loop statement with high arithmetic intensity is extracted using an arithmetic intensity analysis tool such as the ROSE framework (35). Furthermore, loop statements with a large number of loops are also extracted using a profiling tool such as gcov or gprof. OpenCL codes are created using candidate loop statements with a large number of arithmetic intensity and loops. At the time of OpenCL creation, the CPU processing program is divided into the kernel (FPGA) and the host (CPU) in accordance with the OpenCL syntax. For offload candidate loop statements with a large number of arithmetic intensity and loops, our method precompiles the created OpenCL to find a loop statement with high resource efficiency. This is because the resources such as Flip Flop and Lookup Table to be created are known in the middle of compilation, so the loop statements that use a sufficiently small amount of resources are further narrowed down. Since some candidate loop statements remain, our method measures the performance and power consumption using them.

Our method compiles and measures the selected single-loop statement so that it works on the actual FPGA. For a single-loop statement that can be further speeded up, a pattern of the combination is also created and the second measurement is performed. Among the multiple patterns measured in the verification environment, a short-time and low-power pattern is selected as the final solution. For short time and low power, our method uses the similar evaluation value as for GPU (Figure 3).

In summary, after narrowing down the candidate loop statements using the arithmetic intensity, the number of loops, and the resource efficiency, the measurement is performed in the verification environment to increase the evaluation value of the low-power pattern. With them, the speed is increased and the power is reduced automatically.

### 3.3 Automatic offload to mixed environments

I also studied a technology to select an appropriate migration destination while GPU, FPGA, and many-core CPU are mixed as migration destinations.

I propose the following order of verification with three offloads: many-core CPU loop statement offload, GPU loop statement offload, and FPGA loop statement offload. With automatic offload, pattern search is expected to be as quick as possible. Therefore, FPGA verification that takes a long time is the last, and if a pattern that sufficiently satisfies the user requirements is found in the previous stage, FPGA verification will not be performed. There is no big difference in price and verification time between GPU and many-core CPU, but the difference between many-core CPU and normal CPU is smaller than that of GPU with different memory and different devices. Therefore, the verification order is to start with the many-core CPU, and if a pattern that sufficiently satisfies the user requirements is found in the many-core CPU, GPU verification will not be performed.

Here, the previous method is to verify the three migration destinations and automatically select the high-speed migration destination. However, in this paper, not only the short processing time but also the low-power migration destination is automatically selected through the actual measurement in the verification environment. For example, (Processing time)$^{-1/2}$*(Power consumption)$^{-1/2}$ is set to increase evaluation value for short processing time and low power consumption.

As a typical data center cost, the initial cost such as hardware and development cost is 1/3 of the total cost, the operation cost such as power and maintenance is 1/3, and the other cost such as service order is 1/3. In this case, for example, the processing time will be reduced to 1/5, and the initial cost will be reduced if the number of hardware is halved even if the CPU and GPU are combined. Half the power consumption also leads to a reduction in operation cost. However, operation costs have many factors other than electric power, and halving the power consumption does not halve the operation costs. In addition, the hardware price also varies depending on the operator, such as volume discount depending on the number of GPUs and FPGA servers to be installed. Therefore, the evaluation formula needs to be set differently for each business operator.

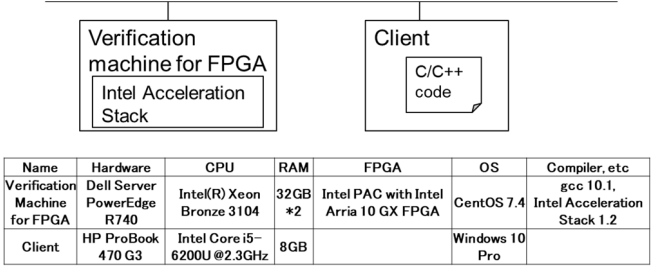

Fig. 4. Experiment environment

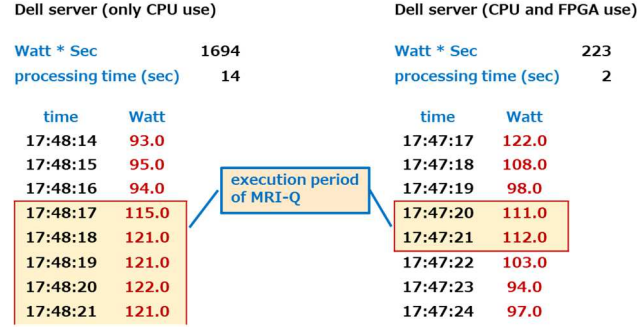

Fig. 5. Power consumption with FPGA offloading (MRI-Q)

In this way, in this paper, the appropriate offload destination is automatically selected in consideration of not only the processing time but also the power consumption.

## 4. Evaluation

The automatic offloading to GPU and FPGA of the loop statement has already been evaluated (33). In this paper, when determining the evaluation value of the measurement pattern on the basis of the implementation of the previous papers, offloading is performed by modifying the implementation to increase the evaluation value of lower power consumption. I will show that the processing time and the power consumption are both reduced by offloading.

### 4.1 Evaluation Condition

(a) Evaluated application

Evaluated application for FPGA offloading is magnetic resonance imaging (MRI) image processing of MRI-Q. This is used in many cases.

MRI-Q (36) computes a matrix Q, representing the scanner configuration for calibration, used in 3D MRI reconstruction algorithms in non-Cartesian space. In an IoT environment, image processing is often necessary for automatic monitoring from camera videos, and performance enhancements are requested in many cases. During application performance measurement, MRI-Q executes 3D MRI image processing to measure processing time using 64*64*64 size sample data. MRI-Q is an application written by C language. It is processed on CPU on the basis of C language logic and processed on FPGA on the basis of OpenCL logic converted from C language codes.

(b) Evaluation method

In the experiment, we enter the code of the evaluation application and the implementation tries to offload the loop statement recognized by an analysis library such as Clang (37) to FPGA. For FPGA offloading, the arithmetic intensity and other values are used to narrow down the measurement patterns to four. At the time of trial, the processing time and power consumption are measured. For the finally determined offload pattern, the time change of the power consumption is acquired, and the improvement of the power consumption compared with the case where all the processing is performed by the normal CPU without offload is confirmed.

Number of processable loop statements. 16 for MRI-Q.

Evaluation value: (Processing time)$^{-1/2}$*(Power consumption)$^{-1/2}$. When processing time and power consumption become smaller, the evaluation value becomes larger. If the performance measurement does not complete in 3 minutes, a timeout is issued, and processing time is set to 1,000 seconds to calculate evaluation value.

(c) Experiment environment

I used physical machines with Intel PAC with an Intel Arria10 GX FPGA for offloading verification and Intel Acceleration Stack Version 1.2 (38) for FPGA control. Regarding power consumption, ipmitool (39) of IPMI (Intelligent Platform Management Interface) equipped with Dell PowerEdge R740 acquired power consumption of a whole server. Figure 4 shows the experimental environment and specifications.

### 4.2 Results

Figure 5 shows Watt and time when MRI-Q was offloaded to FPGA. From the figure, the processing time has been shortened from 14 to 2 seconds compared with the case where all CPU processing is performed. It is also found that the power has been changed from about 121

Watt with only CPU to 111 Watt with CPU and FPGA. As a result, Watt*sec changed from 1,690 Watt*sec when processing only the CPU to 223 Watt*sec when offloaded to the FPGA.

As an application expected to be used by many users, I confirmed the speedup and power reduction of MRI-Q for image processing. When offloaded to the FPGA, the Watt of the whole devices with CPU and FPGA is reduced slightly, which is combined with the shortening of the processing time, resulting in a significant reduction in power consumption. The amount of power consumption during GPU offload is not measured this time, but in a GPU and FPGA mixed environment, an appropriate offload destination is selected from the measured performance and power.

## 5. Conclusions

In this paper, I proposed and evaluated an offload method that considers power consumption as an element of environment-adaptive software for operating applications with high performance and low power.

When actually measuring in the verification environment during automatic GPU and FPGA offloading trials, the power consumption is acquired along with the processing time, and the short-time and low-power pattern is made highly suitable to reduce the power for automatic code conversion. When GPU, FPGA, or so on are mixed, automatic selection is performed by trying migration to a single migration destination and selecting a migration destination with low power consumption and high performance. Through the automatic FPGA offloading of MRI-Q, I demonstrated the high performance and low power consumption and verified the effectiveness of the method.

In the future, I will verify the reduction of power consumption with more applications for both FPGA and GPU. In addition, the evaluation formulas for shortening the processing time and reducing the power consumption will be examined with reference to specific examples of the cost structure of the business operator.

## References

(1) A. Putnam, et al., "A reconfigurable fabric for accelerating large-scale datacenter services," ISCA'14, pp.13-24, 2014.
(2) AWS EC2 web site, https://aws.amazon.com/ec2/instance-types/
(3) O. Sefraoui, et al., "OpenStack: toward an open-source solution for cloud computing," International Journal of Computer Applications, Vol.55, 2012.
(4) Y. Yamato, "Automatic system test technology of virtual machine software patch on IaaS cloud," IEEJ Transactions on Electrical and Electronic Engineering, Vol.10, Issue.S1, pp.165-167, Oct. 2015.
(5) Y. Yamato, "Server Structure Proposal and Automatic Verification Technology on IaaS Cloud of Plural Type Servers," International Conference on Internet Studies (NETs2015), July 2015.
(6) Y. Yamato, "Proposal of Optimum Application Deployment Technology for Heterogeneous IaaS Cloud," 2016 6th International Workshop on Computer Science and Engineering (WCSE 2016), pp.34-37, June 2016.
(7) Y. Yamato, "Use case study of HDD-SSD hybrid storage, distributed storage and HDD storage on OpenStack," 19th International Database Engineering & Applications Symposium (IDEAS15), pp.228-229, July 2015.
(8) Y. Yamato, "Automatic verification technology of software patches for user virtual environments on IaaS cloud," Journal of Cloud Computing, Springer, 2015, 4:4, DOI: 10.1186/s13677-015-0028-6, Feb. 2015.
(9) Y. Yamato, "Automatic verification for plural virtual machines patches," 7th International Conference on Ubiquitous and Future Networks (ICUFN 2015), pp.837-838, July 2015.
(10) Y. Yamato, et al., "Fast and Reliable Restoration Method of Virtual Resources on OpenStack," IEEE Transactions on Cloud Computing, DOI: 10.1109/TCC.2015.2481392, Sep. 2015.
(11) M. Hermann, et al., "Design Principles for Industrie 4.0 Scenarios," Rechnische Universitat Dortmund. 2015.
(12) Y. Yamato, et al., "Proposal of Shoplifting Prevention Service Using Image Analysis and ERP Check," IEEJ Transactions on Electrical and Electronic Engineering, Vol.12, Issue.S1, pp.141-145, June 2017.
(13) Y. Yamato, et al., "Proposal of Real Time Predictive Maintenance Platform with 3D Printer for Business Vehicles," International Journal of Information and Electronics Engineering, Vol.6, No.5, pp.289-293, Sep. 2016.
(14) Y. Yamato, "Proposal of Vital Data Analysis Platform using Wearable Sensor," 5th IIAE International Conference on Industrial Application Engineering 2017

(ICIAE2017), pp.138-143, Mar. 2017.
(15) Y. Yamato, et al., "Security Camera Movie and ERP Data Matching System to Prevent Theft," IEEE Consumer Communications and Networking Conference (CCNC 2017), pp.1021-1022, Jan. 2017.
(16) Y. Yamato, et al., "Analyzing Machine Noise for Real Time Maintenance," 2016 8th International Conference on Graphic and Image Processing (ICGIP 2016), Oct. 2016.
(17) Y. Yamato and M. Takemoto, "Method of Service Template Generation on a Service Coordination Framework," 2nd International Symposium on Ubiquitous Computing Systems (UCS 2004), Nov. 2004.
(18) Y. Yamato, "Ubiquitous Service Composition Technology for Ubiquitous Network Environments," IPSJ Journal, Vol.48, No.2, pp.562-577, Feb. 2007.
(19) Y. Yamato, "Experiments of posture estimation on vehicles using wearable acceleration sensors," The 3rd IEEE International Conference on Big Data Security on Cloud (BigDataSecurity 2017), pp.14-17, May 2017.
(20) P. C. Evans and M. Annunziata, "Industrial Internet: Pushing the Boundaries of Minds and Machines," Technical report of General Electric (GE), Nov. 2012.
(21) T. Sterling, et al., "High performance computing : modern systems and practices," Cambridge, MA : Morgan Kaufmann, ISBN 9780124202153, 2018.
(22) J. E. Stone, et al., "OpenCL: A parallel programming standard for heterogeneous computing systems," Computing in science & engineering, Vol.12, pp.66-73, 2010.
(23) J. Sanders and E. Kandrot, "CUDA by example : an introduction to general-purpose GPU programming," Addison-Wesley, 2011
(24) J. Gosling, et al., "The Java language specification, third edition," Addison-Wesley, 2005. ISBN 0-321-24678-0.
(25) Y. Yamato, "Automatic Offloading Method of Loop Statements of Software to FPGA," International Journal of Parallel, Emergent and Distributed Systems, Taylor and Francis, DOI: 10.1080/17445760.2021.1916020, Apr. 2021.
(26) Y. Yamato, "Improvement Proposal of Automatic GPU Offloading Technology," The 8th International Conference on Information and Education Technology (ICIET 2020), pp.242-246, Mar. 2020.
(27) Y. Yamato, et al., "Automatic GPU Offloading Technology for Open IoT Environment," IEEE Internet of Things Journal, DOI: 10.1109/JIOT.2018.2872545, Sep. 2018.
(28) J. Fung and M. Steve, "Computer vision signal processing on graphics processing units," 2004 IEEE International Conference on Acoustics, Speech, and Signal Processing, Vol. 5, pp.93-96, 2004.
(29) Xilinx SDK web site, https://japan.xilinx.com/html_docs/xilinx2017_4/sdaccel_doc/lyx1504034296578.html
(30) S. Wienke, et al., "OpenACC-first experiences with real-world applications," Euro-Par Parallel Processing, 2012.
(31) M. Wolfe, "Implementing the PGI accelerator model," ACM the 3rd Workshop on General-Purpose Computation on Graphics Processing Units, pp.43-50, Mar. 2010.
(32) E. Su, et al., "Compiler support of the workqueuing execution model for Intel SMP architectures," In Fourth European Workshop on OpenMP, Sep. 2002.
(33) Y. Yamato, "Study of parallel processing area extraction and data transfer number reduction for automatic GPU offloading of IoT applications," Journal of Intelligent Information Systems, Springer, DOI:10.1007/s10844-019-00575-8, Aug. 2019.
(34) J. H. Holland, "Genetic algorithms," Scientific american, Vol.267, No.1, pp.66-73, 1992.
(35) Rose compiler framework web site, http://rosecompiler.org/
(36) MRI-Q website, http://impact.crhc.illinois.edu/parboil/parboil.aspx
(37) Clang website, http://llvm.org/
(38) Intel Acceleration Stack for FPGAs website, https://www.intel.com/content/www/us/en/products/details/fpga/platforms/pac/platform-software.html
(39) Dell IPMI tool web site, https://www.dell.com/downloads/global/power/ps4q04-20040204-murphy.pdf